\documentclass[pdflatex,sn-nature]{sn-jnl}
\usepackage{caption}
\usepackage{graphicx}%
\usepackage{multirow}%
\usepackage{amsmath,amssymb,amsfonts}%
\usepackage{amsthm}%
\usepackage{mathrsfs}%
\usepackage[title]{appendix}%
\usepackage{xcolor}%
\usepackage{textcomp}%
\usepackage{manyfoot}%
\usepackage{booktabs}%
\usepackage{algorithm}%
\usepackage{algorithmicx}%
\usepackage{algpseudocode}%
\usepackage{listings}%
\usepackage{cleveref}

\theoremstyle{thmstyleone}%
\theoremstyle{thmstyletwo}%

\theoremstyle{thmstylethree}%

\begin{document}



\title{A solution to the $S_8$ tension through neutrino–dark matter interactions}

\author*[1]{\fnm{Lei } \sur{Zu}}\email{Lei.Zu@ncbj.gov.pl}

\author*[2]{\fnm{William} \sur{Giarè}}\email{w.giare@sheffield.ac.uk}

\author*[3,4,5]{\fnm{Chi} \sur{ Zhang}}\email{chizhang@pmo.ac.cn}

\author*[2]{\fnm{Eleonora } \sur{Di Valentino}}\email{e.divalentino@sheffield.ac.uk}

\author*[3,4]{\fnm{ Yue-Lin Sming } \sur{Tsai}}\email{smingtsai@pmo.ac.cn}

\author*[1,6]{\fnm{ Sebastian  } \sur{Trojanowski}}\email{Sebastian.Trojanowski@ncbj.gov.pl}

\affil[1]{\orgname{National Centre for Nuclear Research}, \orgaddress{\street{ ul. Pasteura 7}, \city{Warsaw}, \postcode{02-093},  \country{Poland}}}

\affil[2]{\orgdiv{School of Mathematical and Physical Sciences}, \orgname{University of Sheffield}, \orgaddress{\street{Hounsfield Road}, \city{Sheffield}, \postcode{S3 7RH},  \country{United Kingdom}}}

\affil[3]{\orgdiv{Key Laboratory of Dark Matter and Space Astronomy}, \orgname{Purple Mountain Observatory Chinese Academy of Sciences}, \orgaddress{\street{Yuanhua Road}, \city{Nanjing}, \postcode{210033}, \country{China}}}

\affil[4]{\orgdiv{School of Astronomy and Space Science}, \orgname{University of Science and Technology of China}, \orgaddress{\street{Huizhou Dadao},  \city{Hefei}, \postcode{230026}, \country{China}}}

\affil[5]{\orgdiv{International School for Advanced Studies}, \orgname{SISSA}, \orgaddress{\street{Via Bonomea}, \city{Trieste}, \postcode{265}, \country{ Italy}}}

\affil[6]{\orgdiv{Astrocent}, \orgname{Nicolaus Copernicus Astronomical Center Polish Academy of Sciences}, \orgaddress{ul.~Rektorska 4, 00-614}, \city{Warsaw}, \country{Poland}}


\abstract{Neutrinos and dark matter (DM) are two of the least understood components of the Universe, yet both play crucial roles in cosmic evolution. Clues about their fundamental properties may emerge from discrepancies in cosmological measurements across  different epochs of cosmic history. Possible interactions between them could leave distinctive imprints on cosmological observables, offering a rare window into dark sector physics beyond the standard $\Lambda$CDM framework. We present compelling evidence that DM-neutrino interactions can resolve the persistent structure growth parameter discrepancy, $S_8 = \sigma_8\,\sqrt{\Omega_m/0.3}$, between early and late universe observations. By incorporating cosmic shear measurements from current Weak Lensing surveys, we demonstrate that an interaction strength of $u \sim 10^{-4}$ not only provides a coherent explanation for the high-multipole observations from the Atacama Cosmology Telescope (\texttt{ACT}), but also alleviates the $S_8$ discrepancy. Combining early universe constraints with \texttt{DES Y3 cosmic shear} data yields a nearly $3\sigma$ preference for non-zero DM neutrino interactions. This strengthens previous observational claims and provides a clear path toward a significant breakthrough in cosmological research. Our findings challenge the standard $\Lambda$CDM paradigm and highlight the potential of future large-scale structure surveys, which can rigorously test this interaction and unveil the fundamental properties of DM.}




\maketitle

\section{Introduction\label{sec:intro}}

Despite its established role in the Standard Cosmological Model ($\Lambda$CDM), the microscopic nature of dark matter (DM) remains unknown. It is assumed to be cold, i.e., non-relativistic at decoupling, and to interact at most very weakly with baryonic matter~\cite{Roszkowski:2017nbc,Boveia:2022adi,Arcadi:2024ukq}. Cosmology offers a powerful tool for probing the nature of DM, especially in the context of neutrino-DM interactions ($\nu$DM). Both the astrophysical and terrestrial searches for $\nu$DM interactions pose a significant challenge. These interactions can only be indirectly constrained by studying small deviations of neutrino properties from the SM predictions due to new physics~\cite{SajjadAthar:2021prg} or by searching for an excess in the astrophysical neutrino flux caused by DM annihilations in the Galactic center~\cite{Arguelles:2019ouk}. 

Instead, in the early universe, neutrinos contributed significantly to the radiation components, with a number density much higher than baryons. The $\nu$DM interactions induce diffusion-damped dark acoustic oscillations (DAO) in the matter power spectrum~\cite{Mangano:2006mp,Serra:2009uu}. 
Therefore, such sizable couplings lead to deviations from the $\Lambda$CDM model, affecting the cosmic microwave background (CMB) radiation and large-scale structure (LSS) of the universe. Consequently, cosmological observations are highly sensitive to $\nu$DM interactions~\cite{Boehm:2000gq,Boehm:2004th,Serra:2009uu,Mangano:2006mp,Shoemaker:2013tda,Wilkinson:2014ksa,Boehm:2014vja,Bertoni:2014mva,Escudero:2015yka,DiValentino:2017oaw,Olivares-DelCampo:2017feq,Escudero:2018thh,Becker:2020hzj,Paul:2021ewd,Hooper:2021rjc,Dey:2022ini,Brax:2023rrf,Brax:2023tvn,Akita:2023yga,Pal:2023dcs,Dey:2023sxx,Giare:2023qqn,Crumrine:2024sdn,Mosbech:2024wxr,Green:2021gdc,Mosbech:2022uud,Loverde:2022wih}. 

Interestingly, recent analyses of CMB data have revealed a preference for non-zero $\nu$DM couplings, specifically in the high-multipole regime~\cite{Brax:2023rrf,Brax:2023tvn,Giare:2023qqn}, which is consistent with earlier findings in the Lyman-$\alpha$ flux power spectrum~\cite{Hooper:2021rjc}. To further explore these interactions, we test $\nu$DM scenario predictions by incorporating weak lensing (WL) data, a low-redshift observation ($z<3.5$), as a supplement to the high-redshift CMB data. Since the relative impact of $\nu$DM interactions grows at small scales, where nonlinear effects in structure formation become significant, we extend previous analyses by going beyond the linear evolution and including results of $N$-body simulations obtained based on input matter power spectra modified in the presence of $\nu$DM interactions. To make such an analysis feasible, we follow a flexible approach introduced in Ref.~\cite{Zhang:2024mmg}, which allows for conveniently reusing results from past such simulations in various DAO scenarios and global parameter scans. 

WL data are crucial for constraining cosmological parameters, especially for the $S_8$ amplitude defined as $S_8=\sigma_8\sqrt{\Omega_m/0.3}$~\cite{DES:2017myr}, where $\sigma_8$ is the mass dispersion on a scale around $8~h^{-1}\rm{Mpc}$ and $\Omega_m$ is the total matter abundance. In the standard $\Lambda \rm{CDM}$ framework, the $S_8$ value derived from Planck CMB data is larger than the
low-redshift measurements from weak lensing surveys, leading a $2-3 \sigma$ tension~\cite{DES:2021bvc,DES:2021wwk,KiDS:2020suj}. In this work, we revisit this open question within the framework of the $\nu$DM model. By fitting the current three-year Dark Energy Survey (DES) cosmic shear data alone~\cite{DES:2021bvc}, we identify a preferred region for non-zero $\nu$DM interaction strength. Intriguingly, this preferred region is consistent with that favored by ACT data. When combining all early- and late-universe observational data, we find a nearly $3\sigma$ detection of the non-zero $\nu$DM interactions. We show that, for the preferred value of the $\nu$DM interaction strength, both the CMB and WL data lead to consistent fits of the $S_8$ parameter, therefore alleviating the persisting discrepancy.

To further explore the potential of next-generation observations, we generate mock cosmic shear data for upcoming surveys, including the Vera C. Rubin Observatory (formerly known as the Large Synoptic Survey Telescope, LSST)~\cite{LSSTScience:2009jmu}, and the China Space Station Telescope (CSST)~\cite{Gong:2019yxt}. Our results demonstrate that with the improved sensitivity of these future surveys, the favored interaction region will either be robustly confirmed or excluded.

The paper is organized as follows. In \cref{sec:model}, we briefly discuss how $\nu$DM interactions are modeled in our study. \Cref{sec:results} is devoted to presenting the results of our study, and we conclude in \cref{sec:conclusion}. \Cref{sec:data} discusses the implementation of the cosmological datasets in the analysis. The Supplementary  Figure 1 analyzes the expected impact of $\nu$DM scatterings on the matter power spectrum beyond our main assumptions and presents our treatment of the weak lensing data.

\section{Modeling $\nu$DM interactions\label{sec:model}}

In the linear regime of perturbation growth, the shape of the matter power spectrum is determined by solving the Boltzmann equation, which describes the phase-space evolution of the distribution function for various SM species and DM, along with the coupled Einstein and fluid equations~\cite{Kodama:1984ziu,Ma:1995ey}. In the presence of $\nu$DM interactions, additional terms appear in the Boltzmann hierarchy, altering the evolution of perturbations. These interaction terms modify the energy transfer and momentum exchange between DM and neutrinos, leading to distinct imprints on the CMB anisotropies and the matter power spectrum, as detailed in~\cite{Wilkinson:2014ksa,Oldengott:2014qra,Stadler:2019dii,Mosbech:2020ahp}. The corresponding equations are readily solved for cold (non-relativistic) DM species after integrating over momentum. In particular, the equations describing the evolution of the density fluctuations $\delta_\chi$ and the divergence of fluid velocity $\theta_\chi$ are given by
\begin{equation}
\dot{\delta}_\chi = -\theta_\chi + 3\,\dot{\phi},
\end{equation}
\begin{equation}
\dot{\theta}_\chi = k^2\psi - \mathcal{H}\theta_\chi - K_\chi\,\dot{\mu}_\chi\,(\theta_\chi-\theta_\nu),
\label{eq:thetachi}
\end{equation}
where $\phi$ and $\psi$ are scalar potentials appearing in the line element of the perturbed flat Friedmann-Lema\^{i}tre-Robertson-Walker universe, and $\mathcal{H} = \dot{a}/a$ is the Hubble rate.
We define $K_\chi = (1+w_\nu)\,\rho_\nu/\rho_\chi$, where $\rho_\chi$ and $\rho_\nu$ are the DM and neutrino energy densities, respectively, and $w_\nu$ is the neutrino equation of state parameter. For massless neutrinos, one finds $w_\nu = 1/3$ and $K_\chi = (4/3)\,\rho_\nu/\rho_\chi$. From \cref{eq:thetachi}, the evolution of $\theta_\chi$ is modified in the presence of the $\nu$DM interaction term, which is proportional to $\dot{\mu}_\chi$. In the massless neutrino case, this is given by $\dot{\mu}_\chi = a\,n_\chi\sigma_{\nu\textrm{DM}}$, where the cold DM number density is $n_\chi = \rho_\chi/m_\chi$, the $\nu$DM scattering cross section is denoted by $\sigma_{\nu \rm{DM}}$, and $m_{\chi}$ is the DM particle mass.
Therefore, $K_\chi\dot{\mu}_\chi$ depends on the cross section and the inverse of the DM mass. This dependence is commonly parameterized by the dimensionless quantity~\cite{Wilkinson:2014ksa}
\begin{equation}
u_{\nu \rm{DM}}=\frac{\sigma_{\nu \rm{DM}}}{\sigma_{T}}\left(\frac{m_\chi}{100 \,\rm{GeV}}\right)^{-1},
\label{eq:u}
\end{equation}
where $\sigma_{T}$ is the Thomson scattering cross section. The general expression for $\dot{\mu}_\chi$ is more complex for massive neutrinos, but the impact of the $\nu$DM interactions on perturbation evolution can still be effectively described by $u_{\nu \rm{DM}}$ in the cold DM regime. In this work, we consider the most thoroughly studied scenario, in which the neutrinos are massless, and the $\nu$DM interaction cross section is independent of temperature. We refer to \textsl{Supplementary Information} Section 1 for the discussion beyond this approximation and to \cref{sec:data} for details of our cosmological data analysis.

\section{Results\label{sec:results}}

\subsection{Possible evidence of non-zero $\nu$DM interaction strength}

\begin{figure}[t!]
    \centering
   \includegraphics[scale = 1]{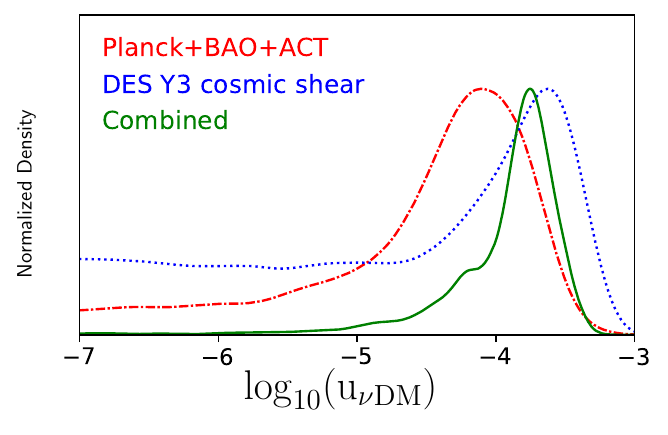}
    \caption{ \textbf{Posterior distributions of the DM–neutrino interaction parameter $\mathbf{u_{\nu\mathrm{\textbf{DM}}}}$.}
    The red dash-dotted(blue dotted) line represents the results obtained using the \texttt{Planck+BAO+ACT}(\texttt{DES Y3 cosmic shear}) likelihood. The combined results for the likelihood \texttt{Planck+BAO+ACT+DES Y3 cosmic shear} are presented as a green solid line.
    }
    \label{Fig:act_u}
\end{figure}

As mentioned above, hints of non-negligible $\nu$DM interaction strength have been found in high-multipole \texttt{ACT}~\cite{Brax:2023rrf,Brax:2023tvn,Giare:2023qqn} and Lyman-$\alpha$~\cite{Hooper:2021rjc} data. For the purpose of this study, we have re-examined the CMB analysis using the \texttt{Planck+BAO+ACT} likelihoods by using the flat priors, shown in Supplementary Table 1. The relevant one-dimensional posterior distribution for $u_{\nu\textrm{DM}}$ is shown as a black solid line in \cref{Fig:act_u}. We observe a preference for nonzero $u_{\nu\textrm{DM}}$ within the $68\%$ credible region, consistent with previous findings, which is driven by the high-$\ell$ \texttt{ACT} data. We note that our results for $u_{\nu\textrm{DM}}$ are slightly shifted to higher values compared to previous works~\cite{Giare:2023qqn}, as we used the full Plik likelihood with a cut at $\ell_{\rm{max}}=650$ rather than the lite version. The preferred parameter regions are reported in \cref{tab:constrain_act}, with a central value of $\rm{log}_{10}\,u_{\nu\textrm{DM}}\simeq -4.24$.

\begin{table}[!h]
    \centering
    \begin{tabular}{|c|c|c|c|c|c|c|c|}
\hline 
Parameter & \multicolumn{1}{c|}{Planck+BAO+ACT} & 

\multicolumn{1}{c|}{+DES Y3 cosmic shear} \\
 
\hline
$100 \Omega_{\rm{b}} h^2$ & $2.235_{-0.014}^{+0.014}$  &$2.247_{-0.014}^{+0.014}$\\
$\Omega_{m} $ &  $0.3060_{-0.0060}^{+0.0060}$ & 
$0.2983^{+0.0048}_{-0.0048}$\\
$100 \theta_s$ & $1.04218_{-0.00049}^{+0.00034}$ &  
$1.04225_{-0.00028}^{+0.00047}$\\
$\ln \left(10^{10} \rm{A_s}\right)$ & $3.036_{-0.015}^{+0.015}$  & $3.029^{+0.016}_{-0.013}$ \\
$n_s$ &  $0.9728_{-0.0047}^{+0.0047}$ &$0.9742^{+0.0046}_{-0.0046}$ \\
$\tau_{\text {reio }}$ &  $0.0487_{-0.0081}^{+0.0069}$ &  $0.0484_{-0.0070}^{+0.0088}$\\
$\log_{10}u_{\nu\rm{DM}}$ & $-4.24_{-0.71}^{+0.56} $ &$-3.70_{-0.34}^{+0.21}$\\
$S_8$ & $0.811_{-0.017}^{+0.024}$ & $0.766_{-0.020}^{+0.024}$\\
\hline
\end{tabular}
    \caption{\textbf{68\% credible intervals for the cosmological parameters.} The $68\%$ credible regions for the cosmological parameters, obtained using the \texttt{Planck+BAO+ACT} and \texttt{Planck+BAO+ACT+DES Y3 cosmic shear} likelihoods. } 
    \label{tab:constrain_act}
\end{table}

We now examine this anomaly using cosmic shear data. The posterior distribution obtained when fitting the \texttt{DES Y3 cosmic shear} likelihood only is shown with the blue dotted line in \cref{Fig:act_u}. 
In this case, we also find a mild preference toward nonzero $u_{\nu\rm{DM}}$, albeit with a lower statistical significance ($<1\sigma$). We discuss details of our cosmic shear analysis in \cref{app:WL} and further illustrate in Supplementary Figures 2 and 3. To test the interplay between the different datasets with regards to this anomaly, we next incorporate a combination of both early- and late-universe observational data, \texttt{Planck+BAO+ACT+DES Y3 cosmic shear}. By employing the combined dataset, we find robust evidence for a nonzero $\nu$DM interaction strength at nearly $3\sigma$ significance, with $u_{\nu\textrm{DM}} \sim 10^{-4}$. This strengthens the preference found in the \texttt{Planck+BAO+ACT} and \texttt{DES Y3 cosmic shear} data. The corresponding posterior distribution for $u_{\nu\rm{DM}}$ is shown with the green solid line in \cref{Fig:act_u}, and the relevant parameter ranges are given in \cref{tab:constrain_act}. The relevant $\chi^2$ values and triangle posterior distributions of the model parameters are presented in the Supplementary Table 2 and Supplementary Figure 4.

\begin{figure}[t!]
\centering
\includegraphics[scale = 0.48]{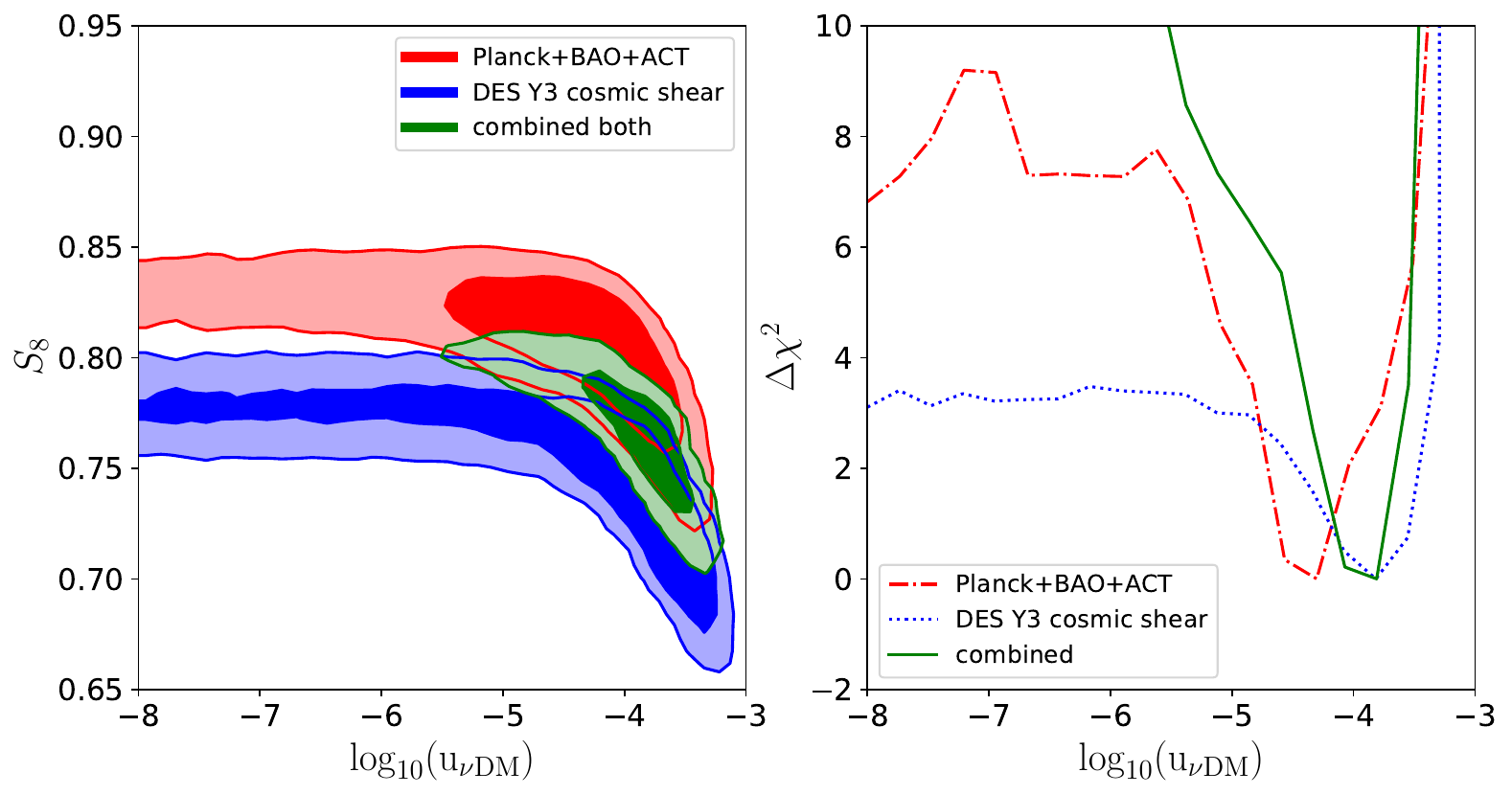}
\caption{\textbf{Profile likelihood distribution and marginalized 2D posterior distribution in the ($\mathbf{S_8}$, $\mathbf{u_{\nu\textrm{\textbf{DM}}}}$) plane.} Left: Marginalized 2D posterior distribution in the ($S_8$, $u_{\nu\textrm{DM}}$) plane. The red and blue contours show the results for the \texttt{Planck+BAO+ACT} and \texttt{DES Y3 cosmic shear} datasets, respectively. The green region represents the results for the combined \texttt{Planck+BAO+ACT+DES Y3 cosmic shear} dataset. Right: The $\Delta\chi^2$ with the parameter $u_{\nu\rm{DM}}$ obtained using the profile likelihood method from the analysis of \texttt{Planck+BAO+ACT}, \texttt{DES Y3 cosmic shear} and combined datasets. }
\label{Fig:s8}
\end{figure}

For a complementary perspective, the right panel of \cref{Fig:s8} also shows the $\Delta\chi^2$ curves obtained from each case using the profile likelihood method. This demonstrates the robustness of the preferred parameter region under different statistical approaches. A detailed breakdown of the $\chi^2$ values for each component is provided in Supplementary Table 1.

This result suggests that the $\nu$DM interaction with $u_{\nu\textrm{DM}} \sim 10^{-4}$ is consistently supported by both CMB and WL cosmological data, despite these being independent observational probes spanning different epochs in cosmic history. This convergence highlights the potential of $\nu$DM interactions as a compelling extension beyond the standard $\Lambda \rm{CDM}$ paradigm, offering new insights into the fundamental nature of DM and its role in cosmic evolution.

It is important to note that interpreting the observed preference in terms of a constant $u_{\nu\textrm{DM}}$ parameter could face challenges from small-scale observations, such as the Lyman-$\alpha$ forest~\cite{Hooper:2021rjc}, dwarf galaxy counts~\cite{Crumrine:2024sdn}, and galaxy luminosity functions~\cite{Mosbech:2022nkk}. We note, however, that these small-scale probes are subject to significant astrophysical uncertainties, particularly those arising from baryonic feedback processes. Such effects can alter the interpretation of structure formation and may introduce non-negligible, model-dependent systematics into the derived constraints on $\nu$DM interactions.

More importantly, the cosmological observables used in our analysis – the CMB and cosmic shear – probe perturbations at different scales and epochs compared to those examined by the Lyman-$\alpha$ forest or galaxy luminosity functions. This apparent tension could be alleviated if the $\nu$DM interaction is not constant but instead varies with redshift, for example, through an energy-dependent scattering cross section as motivated by specific particle physics models~\cite{Trojanowski:2025oro}. Therefore, the working assumption of a constant $\nu$DM cross section applied in our study should be considered a useful and widely adopted phenomenological approximation. It allows for capturing the essential preference in the considered datasets, while acknowledging that further investigations into specific $\nu$DM model implementations should follow.

\subsection{$S_8$ discrepancy}

As mentioned in \cref{sec:intro}, suppression of perturbations at scales probed by WL surveys has an important effect on the matter clustering parameter $S_8 = \sigma_8\,\sqrt{\Omega_m/0.3}$. It has been shown that the persisting tension between the CMB and WL estimates of this parameter, known as the $S_8$ discrepancy (cf. Refs~\cite{DiValentino:2020vvd,Perivolaropoulos:2021jda,Abdalla:2022yfr} for review), can be alleviated by $\nu$DM interactions~\cite{DiValentino:2017oaw}. We revisit this possibility by consistently including the entire \texttt{Planck+BAO+ACT+DES Y3 cosmic shear} dataset in the analysis.

The results of the marginalized 2D posterior distribution in the ($S_8$, $u_{\nu\textrm{DM}}$) plane are presented in the left panel \cref{Fig:s8}. In the plot, we present the results obtained separately for \texttt{Planck+BAO+ACT} and \texttt{DES Y3 cosmic shear} data and the combined analysis. As can be seen, for small values of $u_{\nu\textrm{DM}} < 10^{-6}$, the impact of $\nu$DM interactions is negligible at perturbation scales characteristic to $S_8$. In this case, the $\Lambda$CDM regime is effectively recovered for both the CMB and WL data, and the $2\sigma$ regions obtained for the early and late universe datasets show no overlap. Hence, the $S_8$ discrepancy persists, as expected. 

However, for larger values of $u_{\nu\textrm{DM}}$, the data are consistent with lower values of $S_8$, leading to a better agreement between CMB and WL observations. When combining datasets (\texttt{Planck+BAO+ACT+DES Y3 cosmic shear}), we resolve the $S_8$ tension, as shown with green shaded regions in the plot. Remarkably, the value of the $\nu$DM interaction strength required for this, $u_{\nu\textrm{DM}} \sim 10^{-4}$, corresponds to the previously reported $3\sigma$ evidence found in the combined dataset, cf. also \cref{tab:constrain_act}.

\subsection{Future prospects}

\begin{figure}[t!]
    \centering
   \includegraphics[scale = 0.8]{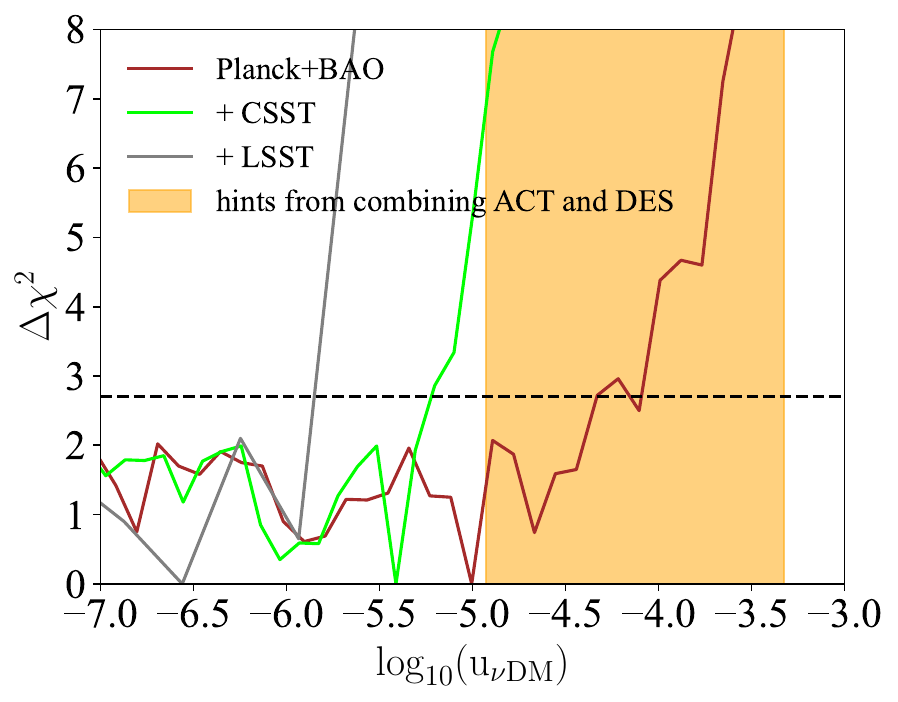}
    \caption{\textbf{Forecasted constraints on the DM–neutrino interaction strength from future weak-lensing surveys.} Variation of $\Delta \chi^2$ with the parameter $u_{\nu\textrm{DM}}$ obtained using the profile likelihood method~\cite{Herold:2024enb}. The brown line represents the \texttt{Planck+BAO} likelihood, while the light green and gray lines additionally include mock cosmic shear data from \texttt{CSST}, and \texttt{LSST}, respectively. The dashed black line indicates $\Delta \chi^2 = 2.71$, corresponding to the $2\sigma$ upper limit. The orange-shaded region indicates the $95\%$ CL preferred range of the $u_{\nu\rm{DM}}$ parameter found in the \texttt{Planck+BAO+ACT+DES Y3 cosmic shear} data.}
    \label{Fig:likelihood}
\end{figure}

The above results highlight the potential of $\nu$DM interactions to address persisting discrepancies in cosmological data that will be decisively studied in next-generation cosmological surveys. To investigate the relevant prospects of future WL observations, we additionally perform MCMC scans, combining the mock data with the \texttt{Planck+BAO+CSST} and \texttt{Planck+BAO+LSST} likelihoods. The resulting likelihood profiles, shown in \cref{Fig:likelihood} (light green and gray lines), indicate that these future WL surveys have the potential to constrain the $\nu$DM interaction parameter to $\rm{log_{10}}\,u_{\nu\textrm{DM}} \lesssim -5.3$ (CSST) or even $\rm{log_{10}}\,u_{\nu\textrm{DM}} \lesssim -5.9$ (LSST) at 95\% CL, assuming the peak we observe is a spurious result and the true cosmology is $\Lambda$CDM, i.e., DM is effectively not interacting with neutrinos at redshifts relevant to our analysis. As can be seen, a significant improvement in sensitivity is expected from these next-generation WL surveys. In particular, the preferred parameter region obtained by fitting ACT and DES Y3 cosmic shear data, as indicated with an orange shading in the plot, will be either thoroughly confirmed or excluded by these surveys. When using mock data generated from a $\nu$DM interacting scenario, the $1\sigma$ error bar on $\rm{log_{10}}\,u_{\nu\textrm{DM}}$ is reduced from $\pm 0.55$ (DES) to $\pm 0.08$ (CSST), demonstrating the promising discovery potential of upcoming WL surveys.

\section{Discussion}\label{sec:conclusion}

Cosmological surveys and gravitational lensing observations provide compelling evidence for the existence of dark matter, but they may also offer critical insights into its fundamental nature. Probing possible interactions of DM with neutrinos remains particularly challenging in terrestrial searches (see Refs~\cite{Berryman:2022hds,Batell:2022xau} for reviews), though this challenge can be mitigated by studying the impact of $\nu$DM scatterings on the matter power spectrum in the early universe. Recent studies have uncovered a slight preference for non-negligible $\nu$DM interactions in high-multipole CMB data and Lyman-$\alpha$ forest observations.

In this study, we utilized weak lensing surveys to investigate $\nu$DM interactions further. We computed the matter power spectrum incorporating nonlinear corrections from $N$-body simulations. These simulations evolve perturbations solely through gravity, as the $\nu$DM interactions are expected to decouple at early times and do not affect matter distributions at the scales probed in our study. We employed an emulator that interpolates matter power spectra from predefined simulations to apply $N$-body results in cosmological parameter scans. 

\texttt{DES Y3 cosmic shear} data alone, which are free from galaxy bias, reveal a preference for non-zero DM-neutrino interaction strength. This preference endures even when combining WL data with CMB and BAO datasets. We find that the \texttt{Planck+BAO+ACT+DES Y3 cosmic shear} data favor non-zero $u_{\nu\rm{DM}}$ at nearly the $3\sigma$ level. This preferred interaction strength, $u_{\nu\rm{DM}}\sim 10^{-4}$, can also simultaneously alleviate the persistent $S_8$ tension. The preference for non-zero $u_{\nu\rm{DM}}$ in the data is additionally linked to a broadening of the $\Omega_m$ distribution, as shown in Supplementary Figure 4. While the $\nu$DM scenario permits a larger value for this parameter, the amount of structures does not increase proportionally. This leads to an overall improved cosmological fit.

Although the statistical significance of these anomalies is not yet sufficient to definitively rule out the standard cosmological scenario, the discrepancies across different observables and datasets underscore the importance of further investigation. Future high-precision WL surveys are expected to provide deeper insights into the mass distribution, particularly at small scales, enabling more stringent tests of $\nu$DM interactions—especially in parameter regions suggested by \texttt{ACT} and \texttt{DES Y3 cosmic shear} observations.

The suppression in the matter power spectrum favored by our analysis could potentially arise in other well-motivated extensions of the $\Lambda$CDM model, beyond $\nu$DM interactions. These include DM interactions with baryons~\cite{Dvorkin:2013cea}, photons~\cite{Boehm:2001hm}, or dark radiation~\cite{Cyr-Racine:2013fsa}, as well as models with warm DM (WDM)~\cite{Bode:2000gq} or ultralight (fuzzy) DM~\cite{Hu:2000ke}. While these models can produce qualitatively similar effects on the linear matter power spectrum, $\nu$DM interactions offer distinct advantages. First, this scenario is less constrained by CMB observations compared to DM interactions with other Standard Model species. Second, unlike WDM or fuzzy DM – where the suppression scale is typically fixed by the DM particle mass – the $\nu$DM interaction can exhibit redshift dependence or affect only a fraction of the dark matter, providing greater flexibility to avoid small-scale bounds; cf. discussion in \textsl{Supplementary Information} Section 1. While $\nu$DM interactions provide a particularly compelling possibility of explaining observed deviations, which can also be independently tested in future terrestrial searches, other beyond $\Lambda$CDM models predicting scale-dependent power spectrum suppression could offer alternative explanations that should also be tested, cf., e.g., Refs~\cite{Amon:2022azi,Preston:2023uup}. The standard cosmological model is under growing pressure, but new hints in cosmological data are driving us toward finding a valid extension.

\paragraph{Note added} After the submission of this paper, the KiDS-Legacy survey reported a higher value for the structure growth parameter, $S_8 = 0.815^{+0.016}_{-0.021}$, which is in good agreement with the Planck $\Lambda$CDM prediction. As these data are not yet publicly available, we cannot assess their impact on our results. However, we stress that the preference we find in our data is not driven by the global $S_8$ tension. Our constraints are primarily informed by cosmic shear measurements sensitive to both quasi-nonlinear scales around $(k \sim 1 \ h\mathrm{Mpc}^{-1}$) and the $S_8$ more linear scales (e.g., $(k \lesssim 0.7 \ h \mathrm{Mpc}^{-1}$)). We do expect that if larger values of $S_8$ are consistently favored by the new data and other WL surveys in the future, then smaller -- yet non-vanishing -- values of $u_{\nu\textrm{DM}}$ might be preferred in the fit, in line with previous findings based on CMB data alone~\cite{Brax:2023rrf,Brax:2023tvn,Giare:2023qqn} and Lyman-$\alpha$ observations~\cite{Hooper:2021rjc}. A status review of this discrepancy can be found in~\cite{CosmoVerse:2025txj}.

\section{Methods\label{sec:data}}

\subsection{Cosmological data analysis} 

We use a modified version of the \texttt{CLASS} code to model the evolution of the universe, specifically accounting for $\nu$DM interactions~\cite{Stadler:2019dii,Mosbech:2020ahp}. In our analysis, we vary $u_{\nu\rm{DM}}$. We also include the six $\Lambda$CDM parameters: the baryon $\Omega_{b}$ and dark matter $\Omega_{\textrm{DM}}$ energy densities (assuming all of DM interacts with neutrinos), the amplitude $\rm{A_s}$ and spectral index $n_s$ of primordial scalar perturbations, the optical depth to reionization $\tau_{\textrm{reio}}$, and the angular size of the horizon at the last scattering surface $\theta$. The prior ranges of each parameters are shown in Supplementary Table 1. We  have verified that using a flat linear prior on the parameter $u_{\nu\rm{DM}}$ does not alter our main conclusions. The effective number of relativistic degrees of freedom is fixed to $\Delta N_{\textrm{eff}}= 3.044$. The following cosmological data are included in the likelihood:

\begin{enumerate}
    \item[(i)] The  DES three-year cosmic shear likelihood ~\cite{DES:2021bvc}. We build an emulator to model the nonlinear correction to the matter power spectrum, calibrated on 200 $N$-body simulations generated with  DAO–modified initial conditions.The linear matter power spectrum is first computed using the modified \texttt{CLASS} code, and the emulator is then used to obtain its nonlinear counterpart; cf. Ref.~\cite{Zhang:2024mmg} for further details. The resulting nonlinear power spectrum is used to predict the cosmic shear signal for a given intrinsic-alignment model; see \cref{app:WL} for more details. We refer to this dataset as \texttt{DES Y3 cosmic shear} throughout this work. We account for nonlinear effects of the gravitational potential on the matter power spectrum at small scales, $k \gtrsim 1~h/\mathrm{Mpc}$ using $N$-body simulations. Notably, on top of the cosmic shear data, the DES Y3 dataset contains galaxy clustering and galaxy-galaxy lensing data. These are, however, subject to unknown galaxy bias, which describes the bias arising from using galaxies as tracers of matter~\cite{Dentler:2021zij,DES:2021wwk}. In this article, we focus on the most robust conclusions based solely on cosmic shear. Further details about the analysis of the DES Y3 dataset and the implementation of nonlinear effects in the matter power spectrum are given in \cref{app:WL}.
    
    \item[(ii)] The CMB likelihoods from Planck 2018 Legacy (P18)~\cite{Planck:2019nip}, including high-$\ell$ power spectra (TT, TE, and EE), low-$\ell$ power spectra (TT and EE), and the Planck lensing reconstruction. The official Planck likelihoods are used directly, and their implementation is interfaced through \texttt{Montepython}. We refer to this dataset as \texttt{Planck}.
    
    \item[(iii)] We use the BOSS DR12 BAO likelihood, which combines distance measurements at $z=0.106$, $0.15$, and $0.2$–$0.75$~\cite{Beutler:2011hx,Ross:2014qpa,BOSS:2016wmc}, referred to as \texttt{BAO}. The likelihood is implemented following the public \texttt{SDSS} likelihood module, assuming Gaussian priors on the measured distance ratios.
    
\noindent For comparison, we also included the new BAO likelihood incorporating the BOSS DR16 dataset~\cite{eBOSS:2020hur,eBOSS:2020uxp,eBOSS:2020fvk,eBOSS:2020tmo,eBOSS:2020yzd} in the data analysis. Including the updated BAO dataset yields nearly identical bounds on $u_{\nu\textrm{DM}}$ as those obtained with DR12, demonstrating the robustness of our results to this update. 
    
    \item[(iv)] The full Atacama Cosmology Telescope (ACT) temperature and polarization DR4 likelihood~\cite{ACT:2020frw}. We use \texttt{HMCode}~\cite{Mead:2016zqy} for nonlinear correction to the matter power spectrum. We have verified that the angular power spectra obtained from \texttt{HMCode} are consistent with those from our emulator at percent level. We refer to this dataset as \texttt{ACT}. When combining the Planck and ACT datasets, we applied a conservative cut of $\ell < 650$ on the Planck data to avoid double-counting in the overlapping multipole range. In this way, the combined dataset utilizes the large-scale information from Planck and the small-scale measurements from ACT. We have additionally confirmed with a sample scan that including the ACT CMB lensing DR6 likelihood~\cite{ACT:2023kun,ACT:2023dou,Carron:2022eyg} does not alter our results. The ACT DR6 lensing data were cut at $\ell<800$ for this purpose.

    \item[(v)] To further investigate the potential of future WL observations, we utilize the expected sensitivity of the upcoming CSST and LSST cosmic shear surveys. Using the publicly available code \texttt{CosmoCov}~\cite{Fang:2020vhc,Krause:2016jvl}, we compute the covariance matrix to represent cosmic shear sensitivity, incorporating the CSST and LSST window functions. The fiducial model we used for these forecasts is the $\Lambda \rm{CDM}$ with  \texttt{Planck} cosmological parameters.
\end{enumerate}

\subsection{Weak Lensing \label{app:WL}}

Weak gravitational lensing allows for directly mapping the late-time Large Scale Structure of the universe by statistically analyzing the shape distortions of numerous galaxies induced by foreground matter fields. The comprehensive set of weak lensing measurements, known as \texttt{3x2pt}, consists of three two-point correlation functions with angular separation $\theta$ of galaxy pairs: galaxy clustering $w(\theta)$ (position-position), galaxy-galaxy lensing $\gamma_t(\theta)$ (position-shape), and cosmic shear $\xi_\pm(\theta)$ (shape-shape). The quantity $w(\theta)$ measures the angular clustering of foreground lens galaxies, while $\gamma_t(\theta)$ measures the correlation between the positions of foreground lens galaxies and the shape distortions of background source galaxies at an angular separation $\theta$. Finally, $\xi_\pm(\theta)$ measures cosmic shear, i.e., the correlation between the shape distortions of background source galaxies due to the foreground LSS. Compared to the galaxy-galaxy lensing and galaxy clustering, the cosmic shear is independent of the galaxy bias, which describes the bias arising from using galaxies as tracers of matter~\cite{Dentler:2021zij,DES:2021wwk}. 

Therefore, the analyses based solely on cosmic shear data lead to the most robust conclusions that we present in the following. When analyzing cosmic shear data, nonlinear effects of the gravitational potential play a significant role in the evolution of LSS at small scales, $k \gtrsim 1~h/\mathrm{Mpc}$. We account for these nonlinear effects using $N$-body simulations.

\subsubsection{$N$-body simulation\label{sec:Nbody}}

As discussed above, $\nu$DM interactions primarily affect weak lensing data through dark acoustic oscillations, which modify the initial matter power spectrum used in $N$-body simulations. Following Ref.~\cite{Zhang:2024mmg} (cf. eqs (3.1) and (3.2) therein), we use the modified Boltzmann code \texttt{CLASS}~\cite{Mosbech:2020ahp} to compute the ratio of linear matter power spectra between $\Lambda$CDM and the $\nu$DM scenario. While this ratio depends on $k$, we effectively reduce its dimensionality to two parameters using principal component analysis (PCA)~\cite{Hotelling_1933}. This allows us to construct a two-parameter grid mapping linear to nonlinear matter power spectra, based on 205 $N$-body simulations run with the \texttt{GIZMO} code~\cite{Hopkins:2014qka,Springel:2005mi}. These simulations were initialized with different matter power spectra corresponding to the interacting DM model. For arbitrary values of $u_{\nu DM}$, we map the corresponding linear power spectrum ratio onto this grid and deduce the nonlinear result through interpolation. Finally, we obtain the nonlinear matter power spectrum for the $\nu$DM scenario by multiplying this nonlinear ratio with the nonlinear $\Lambda$CDM power spectrum from \texttt{Halofit}~\cite{Takahashi:2012em}.

Although the original map of nonlinear power spectrum ratios in Ref.~\cite{Zhang:2024mmg} used DM-baryon interactions, a similar procedure applies to the $\nu$DM case. This is because the parameterization of the linear matter power spectrum ratios via PCA is the same, and the $N$-body simulations evolve solely through gravity. To verify this, we performed an $N$-body simulation for the $\nu$DM scenario with $\rm{log}_{10}u_{\nu\rm{DM}} = -4.6$ and compared the resulting nonlinear matter power spectrum with that obtained from the emulator (i.e., by interpolating on the grid). The Comparison is shown in the left panel of Supplementary Figure 2. This plot has been obtained by neglecting the impact of Halofit. We also present the uncertainty of the simulation in the plot. Because our simulations are performed within a finite comoving volume, therefore the large-scale density fluctuations on scales comparable to or larger than the box size are dominated by the cosmic variance. The relative uncertainty scales approximately as  $\Delta P/P \sim 1/\sqrt{N_{modes} } \approx 1/L_{box}^{3/2}$, highlighting that a larger box volume is required to reduce this large-scale variance, where $N_{modes} $ is the number of independent Fourier modes available in a bin centered at wavenumber k and $L_{box}$ is the box size 200$h^{-1}\rm{Mpc}$. The two results agree well, matching within nearly $15\%$ uncertainty for $k\sim 1~h/\textrm{Mpc}$. 

We also show the uncertainty in $\Delta P(k) = P_{N\rm{-body}}(k)-P_{em}(k)$ in the right panel of Supplementary Figure 2. These results were obtained for the best-fit point in our analysis and for a similar scenario with the same cosmological parameters, except for a lower $\nu$DM interaction strength of $u_{\nu\textrm{DM}}= 10^{-5}$. As can be seen, both results correspond to qualitatively distinct behaviors. This indicates a lack of systematic bias in the emulator results.

To quantify the impact of these differences, we computed the corresponding $\chi^2$ values using the \texttt{DES Y3 cosmic shear} likelihood. For the best-fit point, we found $\chi^2_{\rm emu} = 240.6$ and $\chi^2_{N\text{-}body} = 239.1$ for the most compatible $N$-body simulation result within the uncertainty bars. This difference of $\Delta \chi^2 \approx 1.5$ is well within the statistical uncertainties of the dataset and significantly smaller than the discrepancy introduced by using more approximate non-linear tools, such as $\chi^2_{\rm HMCode} = 399.8$ and $\chi^2_{\rm Halofit} = 8246.6$. Therefore, while the emulator does introduce a modeling uncertainty, it provides a substantially more accurate and reliable non-linear correction compared to \texttt{HMCode} or \texttt{Halofit}, while maintaining the flexibility needed for MCMC scans.

We note, however, that a stronger bias might be introduced by employing the \texttt{Halofit} non-linear power spectrum (obtained based on \texttt{Gadget-2}), when accounting for the impact of variations in other cosmological parameters, as discussed above and in Ref.~\cite{Zhang:2024mmg}. In particular, for the specific points in the parameter space tested in Supplementary Figure 2, we have found a systematic bias between \texttt{GIZMO} and \texttt{Halofit} results of order $\mathcal{O}(10\%)$ for $k\sim 0.1-1$. This may impact the precise value of the best-fit point $u_{\nu\textrm{DM}}$ parameter obtained in the MCMC scan in our analysis, as it appears slightly sensitive to the choice of the baseline simulation results (the difference between $\log{u_{\nu\textrm{DM}}} = -3.7$ and $-4.0$ obtained in additional tests).

\subsubsection{Weak lensing data}

We use the current DES Y3 cosmic shear data in our analysis. This dataset contains the shapes of over $10^8$ source galaxies across an effective area of $4143~\mathrm{deg}^2$. The shape catalog \texttt{METACALIBRATION} used in the DES Y3 analysis is divided into four redshift bins in the redshift range of $0<z<3$~\cite{DES:2021bvc}. Following Ref.~\cite{DES:2021bvc}, we mask small angular scales to reduce uncertainties from baryonic effects. We also utilize cosmic shear mock data from CSST and LSST for future forecasts. The redshift distributions for CSST and DES are different; thus, the CSST mask may not precisely reflect real conditions. However, the capabilities of the future telescope to detect distant galaxies make this mask a conservative estimate. For LSST, we present results with the masking scale set at $l < l_{\rm{max}} = 3000$ following Refs.~\cite{Fang:2019xat,LSSTDarkEnergyScience:2018jkl}.

Supplementary Figure 3 shows the impact of $\nu$DM interactions on the cosmic shear signal. It illustrates the deviation in the expected cosmic shear signal (4th-4th bin) for two selected values of the $\nu$DM interaction strength, $\log_{10}(u_{\nu\rm{DM}}) = -4.6$ and $-3.3$, along with the DES Y3 data points. Nonlinear corrections are applied, as previously described. Stronger interactions lead to greater suppression of the matter power spectrum, which determines the shape of the blue and red curves. For comparison, the linear results are also shown with dotted lines. The nonlinear effects are significant, enhancing the signal at small angular scales. This effect is opposite to that of $\nu$DM interactions; the signal enhancement due to nonlinear effects is partially offset by increasing $u_{\nu\rm{DM}}$.

By treating neutrinos as massless in our analysis, we neglect the impact of their gravitational potential in the $N$-body simulations. We stress that for a total neutrino mass $\sum{m_{\nu}} = 0.06~\textrm{eV}$, the resulting suppression of the nonlinear matter power spectrum is expected to be less than $5\%$ at $k \sim 1~\textrm{h/Mpc}$~\cite{Adamek:2017uiq}. This effect is smaller than the uncertainty introduced by our emulator.

It is worth noting that increasing the neutrino mass beyond this limit can impact the matter power spectrum, particularly at late times and on small scales, and could even affect the inferred value of $S_8$~\cite{Bird:2011rb}. Moreover, since $\sum{m_{\nu}}$ is negatively correlated with $H_0$~\cite{Planck:2018vyg,Vagnozzi:2017ovm}, its inclusion could shift the preferred $H_0$ value downward, potentially exacerbating the $H_0$ tension. However, since the cosmological upper bound on the neutrino mass is primarily driven by BAO data – and these constraints are becoming increasingly stringent~\cite{eBOSS:2020tmo,Jiang:2024viw,Wang:2024hen,DESI:2024mwx} (if not favoring a negative mass~\cite{Green:2024xbb,Elbers:2024sha}) – we adopt a massless neutrino approximation for simplicity.

While our main results are derived from \texttt{cosmic shear} data, we also analyzed the full \texttt{DES Y3 $3 \times 2$pt} dataset for completeness. This analysis also indicates a preference for a non-vanishing $u_{\nu\textrm{DM}}$, although the statistical significance is reduced from nearly $3\sigma$ to below $2\sigma$. In this case, the peak of the posterior distribution for the neutrino interaction parameter is also shifted, favoring a lower interaction strength of $\log_{10}{u_{\nu\textrm{DM}}} = -4.60^{+0.55}_{-3.17}$. We attribute this discrepancy to the limitations of applying a $\Lambda$CDM-based galaxy bias model within our interacting dark sector scenario. Therefore, a more robust and model-compatible treatment of galaxy bias is necessary to draw stronger conclusions from the full \texttt{$3 \times 2$pt} dataset.

\backmatter

\bmhead{Data Availability}

The data used in this study are publicly available from the corresponding survey archives.

The Planck 2018 Legacy Release data can be accessed via the ESA Planck Legacy Archive: \url{https://www.cosmos.esa.int/web/planck/pla}.

The DES Y3 weak lensing and shear catalogues are available from the Dark Energy Survey Data Release Portal:

   The shape catalogs: \url{https://des.ncsa.illinois.edu/releases/y3a2/Y3key-catalogs}

   The cosmic shear data products: \url{https://des.ncsa.illinois.edu/releases/y3a2/Y3key-products}

The ACT DR4 temperature and polarization power spectra are provided by the NASA LAMBDA archive:
\url{https://lambda.gsfc.nasa.gov/product/act/act_dr4_maps_info.html}

The BAO measurements are taken from the BOSS DR12 and eBOSS DR16 galaxy catalogues, accessible from the SDSS Science Archive Server:
\url{https://www.sdss4.org/science/final-bao-and-rsd-measurements-table/}

\bmhead{Acknowledgements}

We would like to thank the anonymous Referees for their useful remarks, which helped to improve our manuscript. This work is supported by the National Key Research and Development Program of China (No. 2022YFF0503304), the China Manned Space Program (No. CMS-CSST-2025-A03), and the Project for Young Scientists in Basic Research of the Chinese Academy of Sciences (No. YSBR-092). CZ is supported by the China Scholarship Council for 1 year study at SISSA. LZ is supported by the NAWA Ulam fellowship (No. BPN/ULM/2023/1/00107/U/00001). LZ and ST are supported by the National Science Centre, Poland (research grant No. 2021/42/E/ST2/00031). ST is also partially supported by Teaming for Excellence grant Astrocent Plus (GA: 101137080) funded by the European Union. EDV acknowledges support from the Royal Society through a Royal Society Dorothy Hodgkin Research Fellowship. W.G. is supported by the Lancaster–Sheffield Consortium for Fundamental Physics under STFC grant: ST/X000621/1. This article is based upon work from the COST Action CA21136 ``Addressing observational tensions in cosmology with systematics and fundamental physics'' (CosmoVerse), supported by COST (European Cooperation in Science and Technology).

\bmhead{Author contributions}

L.Z. conducted the cosmological simulations for this paper, with assistance from W.G. in their implementation. C.Z. was responsible for the N-body simulations, and S.T. wrote the first draft of the manuscript. E.D.V. and Y.-L.S.T. contributed to defining the project's scope and direction and provided insightful advice on interpreting the results. All authors participated in discussions and contributed to the preparation of the final draft.

\bmhead{Competing interests}

The authors declare no competing interests.




\section*{Supplementary Information\label{app:matterpowerspectra}}

Our main discussion focused on the most thoroughly studied scenario, which features a constant $\nu$DM cross-section, i.e., one independent of the early Universe's temperature, and single-component DM. However, relaxing these approximations allows a more straightforward reconciliation of the $\nu$DM scenario with data across various perturbation scales. We will now discuss this in more detail, along with the corresponding impact of these interactions on the matter power spectrum.

\begin{figure}[t!]
\centering
\includegraphics[scale = 0.53]{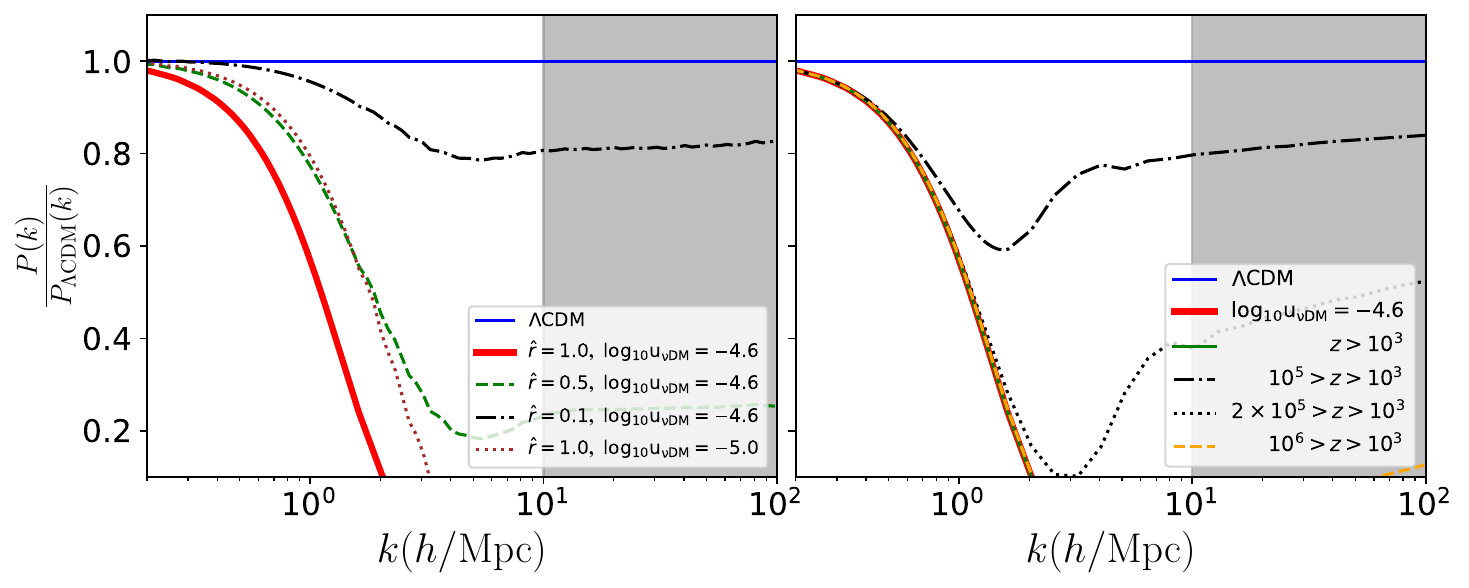}
\caption{\textbf{Impact of DM–neutrino interactions on the linear matter power spectrum.} The ratio of the linear matter power spectrum in the interacting $\nu$DM scenario to that in $\Lambda$CDM at $z=0$. Cosmological parameters are set to $h=0.68$, $\Omega_{\rm{b}} h^2=0.0223$, $\Omega_{\textrm{DM}} h^2=0.120$, $\rm{A}_s=2.215\times 10^{-9}$, $n_s=0.97$, and $\tau_{\rm reio}=0.053$. The topmost solid blue line corresponds to the standard $\Lambda$CDM model. 
The left panel shows results for different fractions of interacting DM, $\hat{r}$. The red solid line corresponds to $\hat{r}=1$ and $\rm{log_{10}}u_{\nu\rm{DM}} = -4.6$. The green dashed and black dash-dotted lines correspond to scenarios with $\hat{r} = 0.5$ and $0.1$, respectively, with the same value of $u_{\nu\rm{DM}}$. For comparison, the brown dotted line shows the case where $\rm{log_{10}}u_{\nu\rm{DM}} = -5$ and $\hat{r} = 1$.
The right panel shows results for scenarios where DM interacts with neutrinos over certain redshift ranges. The red solid line is the same as in the left panel, with the interaction described by a fixed parameter, $\rm{log_{10}}u_{\nu\rm{DM}} = -4.6$, constant across all redshifts. The green line depicts the case where the same value of $u_{\nu\rm{DM}}$ applies, but only for high redshifts $z > 10^3$. The black dotted, dash-dotted and orange long-dashed lines correspond to cases with a non-zero value of $u_{\nu\rm{DM}}$ only for the limited redshift ranges $2\times 10^5 > z > 10^3$, $10^5 > z > 10^3$ and $10^6 > z > 10^3$,  respectively. 
The gray-shaded region indicates scales not used in our weak lensing analysis.
}
\label{Fig:pk_ratio}
\end{figure}

A sample of such an impact relative to $\Lambda$CDM is shown in \cref{Fig:pk_ratio}. Interactions between neutrinos and DM suppress the matter power spectrum at small scales by altering the mass distribution. This is shown with the red solid line, obtained for a benchmark value of $\rm{log_{10}}\,u_{\nu\rm{DM}} = -4.6$. The suppression is substantial (tens of percent) at scales of $k \sim 1~h/\textrm{Mpc}$, which are probed by WL data~\cite{DES:2021wwk}, and it grows even larger at smaller scales.

The left panel also shows results for scenarios where only a fraction of DM interacts with neutrinos, as described by the parameter $\hat{r}= \Omega_{\nu\rm{DM}}/\Omega_{\rm{DM}}$, where $\Omega_{\nu\rm{DM}}$ is the relic abundance of the interacting DM component and $\Omega_{\rm{DM}}$ corresponds to the total DM relic density. We consider three different fractions of interacting DM: $\hat{r} = 1$, $0.5$, and $0.1$. A smaller $\nu$DM fraction leads to a milder impact on the power spectrum, with noticeable effects emerging at larger values of $k$. In particular, the lines representing the relative spectra for $\hat{r}<1$ flatten at $k\gtrsim \textrm{a few}~h/\rm{Mpc}$; see also Ref.~\cite{Zu:2023rmc} for a similar discussion. In comparison, we also show the spectrum obtained for $\hat{r}=1$ but for a lower value of $\rm{log_{10}}\,u_{\nu\rm{DM}} = -5$. Varying $u_{\nu\rm{DM}}$ causes a large-scale suppression comparable to that caused by varying $\hat{r}$, though they still differ significantly at small scales. The presence of a non-interacting cold DM component significantly weakens the impact of $\nu$DM scatterings on the spectrum observed at high $k$.

The matter power spectrum is also non-trivially affected if $\nu$DM interactions effectively decouple outside a limited redshift range. We first note that the behavior of the cross section at low redshift does not affect perturbations at scales relevant to our discussion, as they are effectively set at redshifts $z \gtrsim 10^3$, i.e., before recombination. This is because neutrinos and DM effectively decouple at low redshifts. We illustrate this in the right panel of \cref{Fig:pk_ratio}. The green dashed line corresponds to the scenario with $\rm{log_{10}}\,u_{\nu\rm{DM}} = -4.6$ at these high redshifts, while negligible values of this parameter are considered at later epochs, i.e., $u_{\nu\rm{DM}} = 0$ for $z < 10^3$. This line assumes that all DM interacts with neutrinos, $\hat{r} = 1$. The resulting matter power spectrum is nearly identical to the red solid line, which assumes a constant value of $u_{\nu\rm{DM}}$ for all redshifts, including low $z$. For the considered value of $u_{\nu\rm{DM}}$, decoupling occurs in the mixed-damping regime~\cite{Stadler:2019dii}. In this case, the DM interaction rate with neutrinos, $\Gamma_{\rm{DM}-\nu}$, decouples as early as $z \sim 10^4$ for a temperature-independent cross-section. The primary impact of $\nu$DM interactions on structure formation is then through the matter power spectrum set at high redshifts. In our analysis, this serves as input to $N$-body simulations that begin evolving perturbations at $z \sim 100$ and neglect direct $\nu$DM scatterings at late times.

Possible late-time neutrino interactions become increasingly important at smaller scales, i.e., in overdense DM regions where the DM density is much higher than the background density. This is especially important at larger $k$, where additional effects like baryonic feedback should be considered. A full treatment of $\nu$DM interactions in $N$-body simulations is left for future work. In this analysis, we cut the WL datasets at $k \lesssim \textrm{a few}~h/\textrm{Mpc}$, below which these effects are not expected to significantly affect our results. This is indicated by the gray-shaded region for $k \gtrsim 10~h/\textrm{Mpc}$. While this limits the predicted sensitivities of the WL data, it allows for a conservative estimate of the capabilities of future WL observations.

At very early times, the impact of $\nu$DM interactions on the matter power spectrum is also limited for the scales of interest. This is illustrated by the dashed orange line in the right panel of \cref{Fig:pk_ratio}. For this line, the matter power spectrum was calculated assuming the aforementioned value of $u_{\nu\rm{DM}}$ only within the limited range of $10^3 < z < 10^6$, with a negligible $\nu$DM coupling strength outside this redshift interval. Specifically, the interactions were disregarded at higher redshifts.

For comparison, the plot also shows the matter power spectra obtained using more stringent upper redshift cuts of $10^3 < z < 2 \times 10^5$ and $10^3 < z< 10^5$ (dotted and dot-dashed black lines, respectively). These spectra show a suppression of up to a few tens of percent at $k \sim \textrm{a few}~h/\textrm{Mpc}$ relative to $\Lambda$CDM, depending on the precise upper redshift cut. However, this suppression becomes less pronounced at smaller scales, where we observe a flattening similar to that seen in the $\hat{r}<1$ scenarios.

This is important because small-scale suppression of the matter power spectrum is expected to impact the distribution of low-mass DM halos and dwarf galaxies around the Milky Way~\cite{Boehm:2014vja,Akita:2023yga,Crumrine:2024sdn}; cf. also bounds obtained based on galaxy luminosity function~\cite{Mosbech:2022nkk} and Lyman-$\alpha$ observations~\cite{Hooper:2021rjc}. The faintest observed dwarf galaxies correspond to wavenumbers of $k \sim 10-100~h/\textrm{Mpc}$, which goes beyond the typical constraining power of WL data. At these small scales, the suppression of the matter power spectrum could surpass that at larger scales for fixed $u_{\nu\rm{DM}}$, potentially leading to strong bounds on the $\nu$DM interaction strength. However, the relevant modes enter the horizon at early times and are primarily affected by $\nu$DM interactions at redshifts of $z \gtrsim 10^6$~\cite{Crumrine:2024sdn}, which may differ from the lower redshifts where the dominant effect on WL and CMB data is expected; cf. recent discussion of such fits for redshift-limited $\nu$DM interactions~\cite{Trojanowski:2025oro}. Further investigation is needed to fully assess the constraining power of dwarf galaxy observations, especially considering potential systematic uncertainties related to their luminosity function and other astrophysical factors.

Neglecting the late- and early-time impacts of $\nu$DM interactions can also be well-grounded in underlying particle physics scenarios. Testing $\nu$DM interactions at different redshifts is equivalent to probing them in different energy regimes. As the universe expands and cools down, the average neutrino energy is given by $\langle E_\nu^2 \rangle = 15\,[\xi(5)/\xi(3)]\,T_\nu^2$, assuming Fermi-Dirac statistics and negligible neutrino masses. Simple $\nu$DM portals typically imply that $\sigma_{\nu\rm{DM}} \propto E_\nu^n \propto T^n_\nu$ with $n = 2$ or $4$, so the cross section decreases with decreasing temperature~\cite{Olivares-DelCampo:2017feq}. However, the energy dependence of the cross section is generally more complex in more complete beyond the Standard Model (BSM) frameworks, cf. Refs.~\cite{GonzalezMacias:2015rxl,Blennow:2019fhy} for further discussion. To avoid violating perturbative partial-wave unitarity beyond the effective field theory regime, the cross section should stop growing at high temperatures. This helps to avoid stringent bounds on $u_{\nu\rm{DM}}$ from the attenuation of high-energy neutrino flux from distant blazars~\cite{Cline:2022qld,Ferrer:2022kei,Cline:2023tkp} and other astrophysical probes, cf., e.g., Refs.~\cite{Farzan:2014gza,Arguelles:2017atb,Pandey:2018wvh,Kelly:2018tyg,Alvey:2019jzx,Choi:2019ixb,Jho:2021rmn,Ghosh:2021vkt,Lin:2022dbl,Lin:2023nsm,Fujiwara:2023lsv,Heston:2024ljf,Lin:2024vzy,Fujiwara:2024qos}. Thus, a characteristic range of energies at which the scattering cross section between neutrinos and DM is maximized might naturally emerge in realistic $\nu$DM scenarios. This corresponds to a limited redshift range in the early universe.

One such example is the $\nu$DM interaction mediated by the sterile neutrino portal, introduced to address small-scale structure tensions in $\Lambda$CDM~\cite{Bertoni:2014mva,Batell:2017rol,Batell:2017cmf}. In this case, depending on the mass splitting between the DM and an additional, heavier dark state in the model, $u_{\nu\rm{DM}}$ can be effectively constant with temperature in a limited range of $z$ before recombination, while decreasing rapidly outside this range at both higher and lower redshifts~\cite{Brax:2023tvn}; cf. also earlier discussion~\cite{Boehm:2003hm}. Redshift-limited enhancements in the interaction parameter $u_{\nu\textrm{DM}}$ can also be obtained in the resonant regime~\cite{Trojanowski:2025oro}, which highlights the importance of probing $\nu$DM interactions using a wide range of cosmological and astrophysical observables. In this work, the assumption of a constant $\nu$DM scattering cross section serves as a simplified phenomenological approximation. This approach allows us to capture the essential impact of $\nu$DM interactions without making the analysis overly model-dependent.


\begin{figure}
    \centering
    \includegraphics[width=0.48\linewidth]{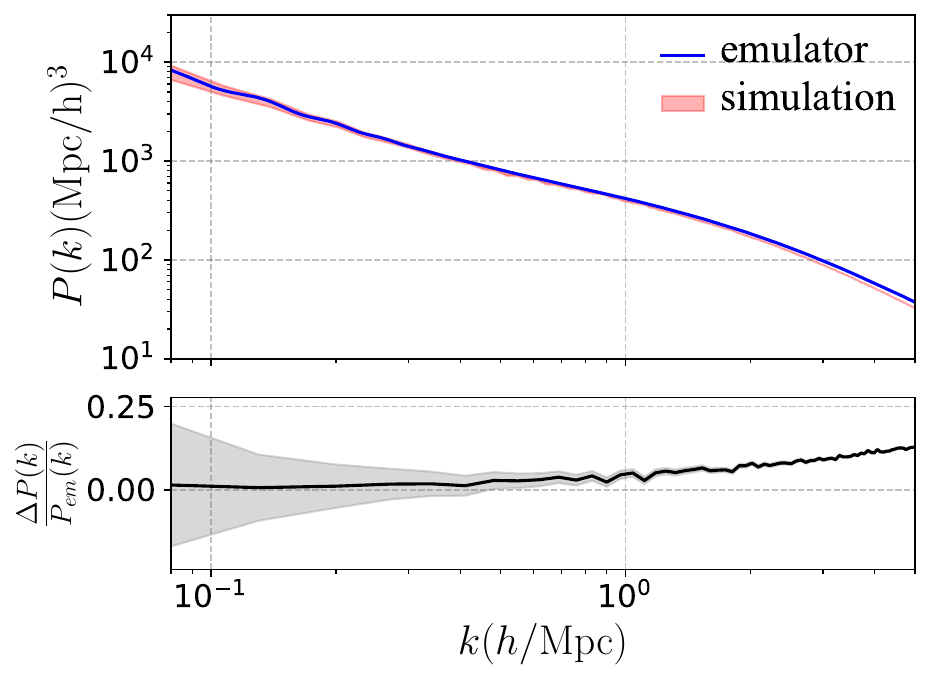}
    \includegraphics[width=0.48\linewidth]{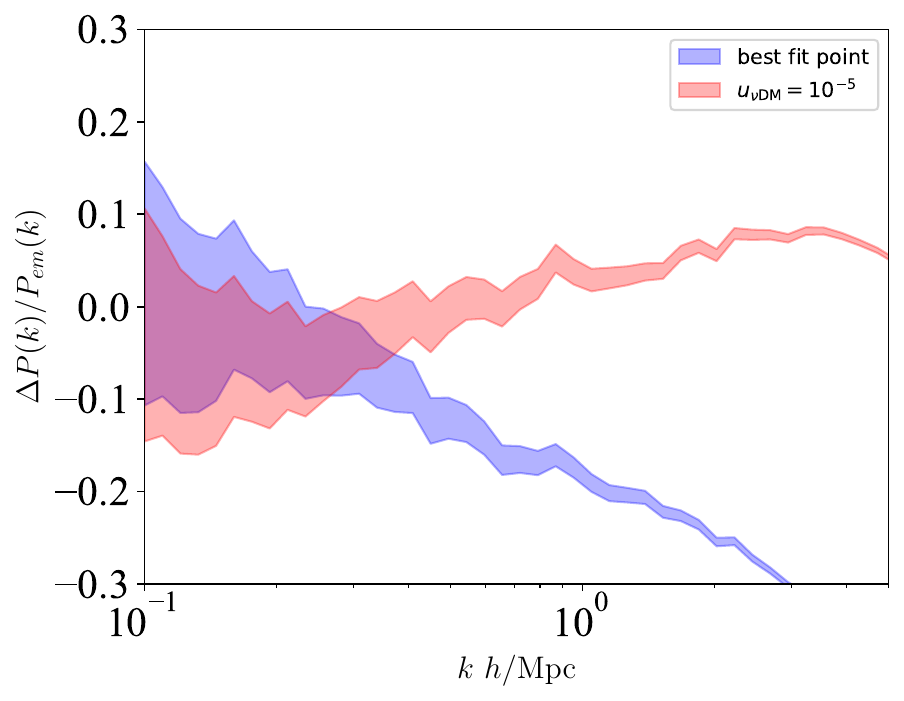}
    \caption{\textbf{Validation of the emulator against $\mathbf{N}$-body simulations.} Left: Comparison of the nonlinear matter power spectrum, $P(k)$, obtained from an emulator (blue line) with that from a full $N$-body simulation (red) at z=0. Cosmological parameters are set to $h=0.68$, $\Omega_{\rm{b}} h^2=0.0223$, $\Omega_{\rm{DM}} h^2=0.120$, $\rm{A_s}=2.215\times 10^{-9}$, $n_s=0.97$, $\tau_{\rm reio}=0.053$, and $\rm{log}_{10}u_{\nu\rm{DM}}=-4.6$. The lower panel shows the percentage difference between the emulator and $N$-body results. The shaded band indicates the statistical uncertainty of the simulations, arising from finite box size (200 $h^{-1}$Mpc) and limited realization sampling. As a result, the power spectrum measurements at small wavenumbers are affected by sample variance. Right: The quantity $\Delta P(k)/P_{em}(k)$, which corresponds to the relative difference between the matter power spectra in $N$-body simulation and the result obtained with the emulator at z=0, as a function of wavenumber $k \ [h/\mathrm{Mpc}]$, shown with the statistical uncertainty of the simulations. The blue band corresponds to the best-fit point in our analysis, while the red band represents the same cosmological parameters with the exception that $u_{\nu\rm{DM}}=10^{-5}$.}
    \label{fig:pk_simu_emu}
\end{figure}

\begin{figure}[t!]
\centering
\includegraphics[scale = 0.8]{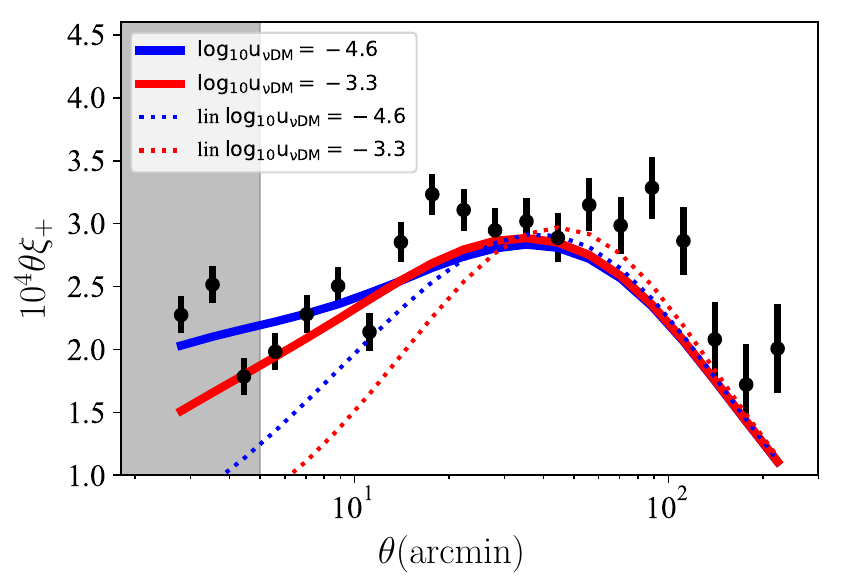}
\caption{\textbf{Comparison of DES Y3 cosmic shear measurements with predictions from the $\nu$DM interaction model.} Cosmic shear signal in the 4th-4th redshift bin of the DES Y3 data alongside predictions of the $\nu$DM scenario. The DES Y3 data are shown as mean $\pm$ standard error of the mean (SEM), with the SEM derived from the diagonal elements of the covariance matrix. The solid blue (red) line shows the results for $\rm{log}_{10}u_{\nu\rm{DM}}=-4.6$ ($-3.3$). Nonlinear corrections are included using the emulator discussed in Section 5.1.1. For Comparison, the linear results are shown as dotted lines. The gray region is masked to avoid uncertainties related to baryonic feedback.}
\label{Fig:cosmic_shear}
\end{figure}

\begin{table}
    \centering
    \begin{tabular}{|c|c|c|c|c|c|c|c|}
\hline
\hline
$100 \Omega_{\rm{b}} h^2$  & [2.147, 2.327] \\
$\Omega_{\textrm{DM}} h^2$ &   [0 , 0.2] \\
$100 \theta_s$ &  [1.0393,  1.0429] \\
$\ln \left(10^{10} \rm{A_s}\right)$  & [2.9547,  3.1347] \\
$n_s$ &   [0.9407, 0.9911]  \\
$\tau_{\text {reio }}$ &    [0.01,  0.7]  \\
$\log_{10}u_{\nu\rm{DM}}$  & [-8.0, -3.0] \\
\hline
\hline

\end{tabular}
\captionsetup{width=1\textwidth} 
    \caption{\textbf{Cosmological parameters and their prior ranges used in the MCMC analysis.}  
    All parameters are assigned flat (uniform) priors within the specified ranges.
}
    \label{tab:prior}
\end{table}

\begin{table}
    \centering
    \begin{tabular}{|c|c|c|c|c|c|c|c|}
    \hline
& \multicolumn{1}{c|}{Planck+BAO+ACT} & 

\multicolumn{1}{c|}{DES Y3 cosmic shear} & Combined \\
 
\hline
Planck+BAO & 1216 & - & 1221 \\
ACT & 289 & -& 290 \\
DES cosmic shear & - &234 & 236\\
\hline
Total & 1505 &234 &1747\\
\hline
    
    \end{tabular}
    \captionsetup{width=1\textwidth}
    \caption{\textbf{Best-fit $\chi^2$ values.} Best-fit $\chi^2$ values for each dataset, as well as the total, from fits to different combinations of experiments.  } 
    \label{tab:chi2}
\end{table}

\begin{figure}[p]
    \centering
    \includegraphics[scale = 0.3]{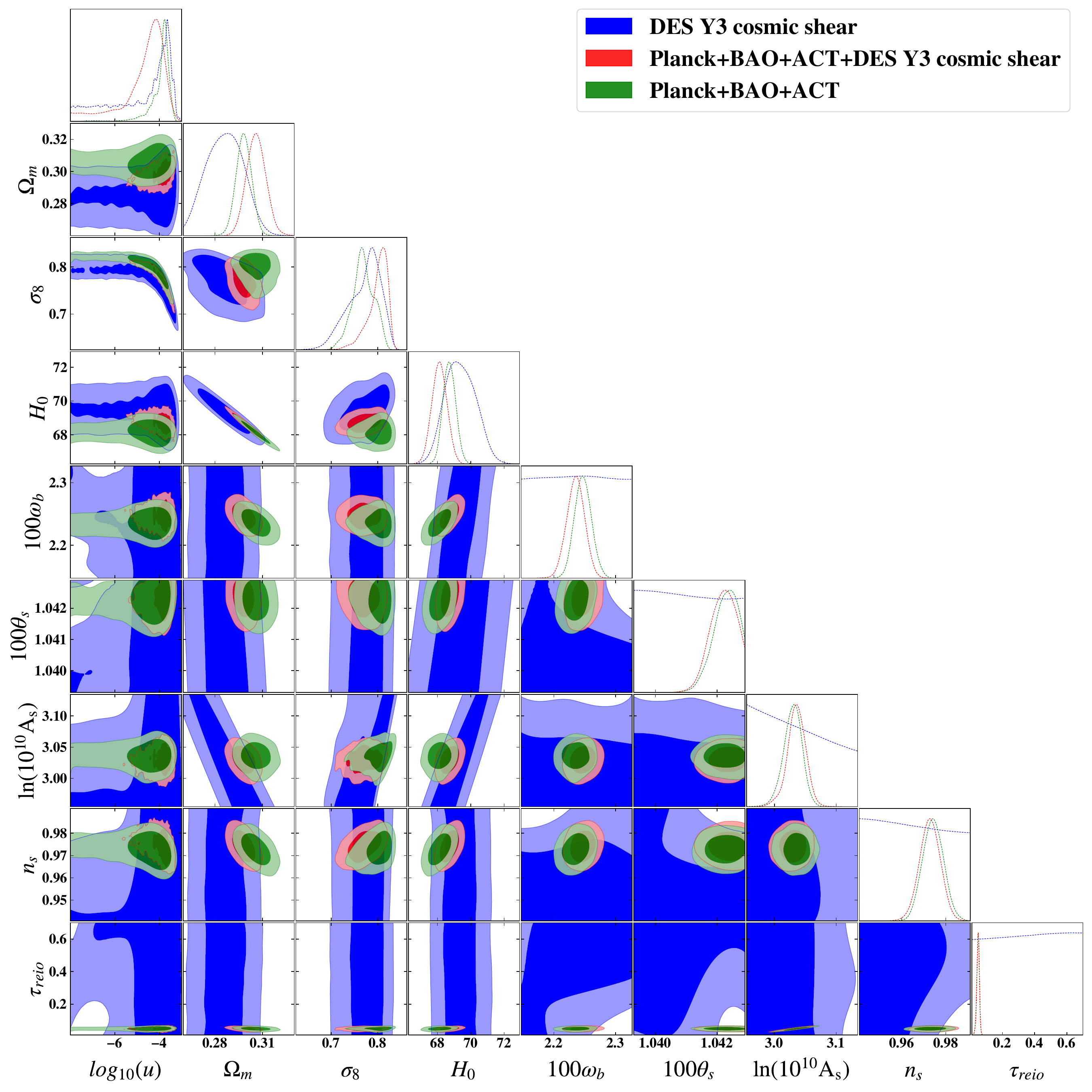}
    \caption{The triangle posterior distribution of MCMC results in the $\nu$DM interaction model with the dataset \texttt{Planck+BAO+ACT}(green), \texttt{DES Y3 cosmic shear}(blue) and \texttt{Planck+BAO+ACT+DES Y3 cosmic shear}(red). The plot assumes $100\%$ interacting DM, $\hat{r} = 1$.}
    \label{Fig:triangle_cosmicshear}
\end{figure}
\clearpage

\bibliography{suppl-bibliography}

\end{document}